\documentstyle[preprint,aps,version2]{revtex}
\begin{document}
\draft
\preprint{OCIP/C 94-9}
\preprint{hep-ph/yymmxxx}
\preprint{August 1994}
\begin{title}
Comparison of Discovery Limits \\
for Extra $Z$ Bosons at Future Colliders \\
\end{title}
\author{Stephen Godfrey\footnote{e-mail: godfrey@physics.carleton.ca}}
\begin{instit}
Ottawa-Carleton Institute for Physics \\
Department of Physics, Carleton University, Ottawa CANADA, K1S 5B6
\end{instit}

\begin{abstract}
We study and compare the discovery potential
for heavy neutral gauge bosons ($Z'$) at various $e^+e^-$ and
$p\stackrel{(-)}{p}$ colliders that are planned or have been
proposed.  Typical discovery limits are for the TEVATRON $\sim
1$~TeV,  DI-TEVATRON $\sim 2$~TeV, LHC $\sim 4$~TeV,
LSGNA (a 60~TeV $pp$ collider) $\sim 13$~TeV
while the $e^+e^-$ discovery limits are $2-10\times
\sqrt{s}$ with the large variation reflecting the model dependence of the
limits.
While both types of colliders have comparable discovery
limits the hadron colliders are generally less dependent on the specific
$Z'$ model and provide more robust limits since the signal has
little background.  In contrast, discovery limits for $e^+e^-$
limits are more model dependent and because they are
based on indirect inferences of deviations from standard
model predictions, they are more sensitive to systematic errors.
\end{abstract}

\section{INTRODUCTION}

Extended gauge symmetries and the associated heavy neutral gauge
bosons, $Z'$, are a feature of many extensions of the standard model
such as grand unified theories, Left-Right symmetric models, and
superstring theories.  If a $Z'$ were discovered it would have
important implications for what lies beyond the standard model.
It is therefore a useful excercise to study and compare the discovery
reach for extra gauge bosons at the various facilities that
will operate during the next decade (TEVATRON and LEP200) and future
facilites that are being planned or are under consideration for the
period beyond (various TEVATRON upgrades, LHC, the NLC $e^+e^-$
collider, and LSGNA, a 60~TeV $pp$ collider \cite{haber}).
Such a comparison was made in earlier
papers\cite{capstick,hewett,hewett2}
but since those papers were published
many new facilities have been proposed making it a
useful excercise to update those analysis.
In this brief report we examine and
compare the discovery limits for extra neutral gauge bosons at high
energy $e^+e^-$, and hadron colliders that are being built or
have been proposed.  The collider parameters are listed in fig. 1.
The goal is to compare the relative strengths
and weaknesses of these facilities.

\section{MODELS}

Quite a few models predicting extra gauge bosons exist in the
literature.  We will present
discovery limits for several of these models which, although far
from exhaustive, we feel form a  representative set for the purposes
of comparison.  For the benefit of the reader we briefly describe
the models we have chosen to study.

(i) Effective rank-5 models \cite{er5m} originating from $E_6$ grand unified
theories are conveniently labelled in terms
of the decay chain $E_6 \to SO(10) \times U(1)_\psi \to SU(5)\times
U(1)_\chi \times U(1)_\psi \to SM \times U(1)_{\theta_{E_6}}$.
Thus, the $Z'$ charges are given by linear combinations of the
$U(1)_\chi$ and $U(1)_\psi$ charges resulting in the $Z'$-fermion
couplings:
\begin{equation}
g_{Z^0} ({g_{Z'}\over g_{Z^0}})
(Q_\chi \cos \theta_{E_6} + Q_\psi \sin\theta_{E_6})
\end{equation}
where $\theta_{E_6}$ is a free parameter which lies in the range
$-90^\circ \leq \theta_{E_6} \leq 90^\circ$, $(g_{Z'}/g_{Z^0})^2
\leq {5\over 3}\sin^2\theta_w$ (here we assume the equality),
$c_w=\cos\theta_w$,
and $Q_\psi (Q_\chi ) = [1,1,1]/2\sqrt{6}
\; ([-1, 3, -5]/2\sqrt{10})$ for $[(u,d,u^c,e^c), \; (d^c, \nu e^-), \;
(N^c)]$, the left-handed fermions in the {\bf 10}, $\overline{\bf 5}$,
and {\bf 1} of $SU(5)$ contained in the usual {\bf 16} of $SO(10)$.
Specific models of interest are model $\chi$ ($\theta_{E_6}=0^\circ$)
corresponding to the extra $Z'$ of $SO(10)$, model $\psi$
($\theta_{E_6}=90^\circ$) corresponding to the extra $Z'$ of $E_6$,
and model $\eta$ ($\theta_{E_6}=\arctan -\sqrt{5/3}$)
corresponding to the extra $Z'$ arising in some superstring theories.

(ii) The Left-Right symmetric model (LRM) extends the standard model
gauge group to $SU(2)_L \times SU(2)_R \times U(1)$ \cite{mohapatra}.
The $Z'$-fermion coupling is given by
\begin{equation}
g_{Z^0}
{1\over {\sqrt{\kappa - (1+\kappa)x_W}}} [x_W T_{3L} + \kappa (1-x_W
) T_{3R} -x_W Q]
\end{equation}
with $0.55 \leq \kappa^2 \equiv (g_R/g_L)^2 \leq 1-2$  \cite{chang},
$T_{3L(R)}$ the isospin assignments of the fermions under
$SU(2)_{L(R)}$, $Q$ the fermion electric charge and
$x_W=\sin\theta_W$.  We assume $\kappa=1$ in our analysis which
corresponds to strict left-right symmetry.  Note
that the $T_{3L}$ assignments are the same as in the standard model
while the values of $T_{3R}$ for $u_R, \; d_R, \; e_R, \; \nu_R =
{1\over 2}, \; -{1\over 2}, \; -{1\over 2}, \; {1\over 2}$ and are
zero for left-handed doublets.

(iii) The Alternative Left-Right Symmetric model (ALRM) \cite{ma}
originates from $E_6$ GUT's and is also based on the electroweak
gauge group $SU(2)_L \times SU(2)_R \times U(1)$.  Here the
assignments for $T_{3L(R)}$ differ from that of the usual LRM for
$\nu_{L,R}$, $e_L$, and $d_R$ with
$T_{3L(R)}(\nu_L)= {1\over 2} (-{1\over 2})$,
$T_{3L(R)}(e_L)= -{1\over 2} (-{1\over 2})$,
and $T_{3L(R)}(d_R)=0$.  The LRM and ALRM have idential $u$-quark,
$e_R$, and $d_L$ couplings.

(iv) The ``sequential'' Standard Model (SSM) consists of a $Z'$
which is just a heavy version of the SM $Z^0$ boson with identical
couplings. Although it is not a realistic model
it is often used as a benchmark and for purposes of comparison.

(iv) The Harvard Model (HARV) \cite{georgi} is based on the gauge
group $SU(2)_l \times SU(2)_q \times U(1)_Y$, i.e., left-handed
leptons (quarks) transform as doublets under $SU(2)_l$ ($SU(2)_q$)
and singlets under  $SU(2)_q$ ($SU(2)_l$), and right-handed fields
are singlets under both groups.  The $Z'$-fermion coupling takes the form
\begin{equation}
g_{Z^0} c_w \left( { {T_{3q}\over {\tan\phi}} -\tan\phi T_{3l} }\right)
\end{equation}
where $T_{3q(l)}$ is the third component of $SU(2)_{q(l)}$-isospin,
$c_w=\cos\theta_w$, and $\phi$ is a mixing parameter which lies in
the range $0.22 \leq \sin\phi \leq 0.99$.  We take $\sin\phi=0.5$ in
our calculations.  The $Z'$ is purely left
handed in this model.

There are numerous other models predicting $Z'$'s in the literature
\cite{othermodels} but the subset described
above have properties reasonably representative of the broad class of
models,  at least for the purposes of comparing discovery
limits of high energy colliders.

\section{PRESENT LIMITS}

Before proceeding to future colliders it is useful to list
existing bounds as a benchmark against which to measure future
experiments.  Constraints can be placed on the
existence of $Z'$'s either indirectly from fits to high precision
electroweak data\cite{langacker,ew}
or from direct searches at operating collider facilities\cite{cdf}.

There have been a number of fits to precision data
\cite{langacker,ew}.  We list results from the particle data group
\cite{pdg} in Table I.
These results contain no assumptions on the Higgs sector.

The highest mass limits come from direct searches by the CDF
experiment at the Tevatron\cite{cdf}.  The CDF limits are obtained by looking
for high invariant mass lepton pairs that would result from a $Z'$
being produced via the Drell-Yan mechanism \cite{pp}
and subsequently decay to
lepton pairs, $p\bar{p}\to Z' \to \ell^+\ell^-$.
The most recent CDF 95 \% confidence level results based on ${\cal
L}_{int}=19.6\hbox{pb}^{-1}$ are listed in Table I.

\section{FUTURE LIMITS}

Bounds on extra gauge bosons attainable from low energy neutral
current precision measurements, measurements at the TRISTAN, LEP,
and SLC $e^+e^-$ colliders, as well as at the HERA $ep$ collider
have been surpassed by direct limits obtained at the TEVATRON
$p\bar{p}$ collider or will be from future TEVATRON upgrades. Thus,
we will restrict our results to LEP200, proposed TEVATRON upgrades,
the LHC and LSGNA $pp$ colliders and the NLC high luminosity $e^+e^-$
colliders.

\subsection{Hadron Colliders}

The signal for a $Z'$ at a hadron collider consists of
Drell-Yan production of lepton pairs \cite{pp} with
high invariant mass via
$p \stackrel{\scriptsize (-)}{p} \to Z' \to l^+ l^-$.
The expressions for this process are given in ref. \cite{capstick}.
We obtain the discovery limits for this process based on 10 events
in the $e^+e^- + \mu^+\mu^-$ channels using the EHLQ structure
functions \cite{ehlq}, taking $\alpha=1/128.5$,
$\sin^2\theta_w=0.23$, and including a 1-loop $K$-factor in the $Z'$
production. We include a $t$-quark of mass 174~GeV in the $Z'$ decay
width, and 2-loop QCD radiative corrections and 1-loop QED radiative
corrections in calculating the $Z'$ width.  Using different
quark distribution functions results in a roughly 10\% variation in
the $Z'$ cross sections\cite{rizzo} with the subsequent change in
discovery limits.  We note that including
realistic detector efficiencies would lower these limits.

In our calculations we assumed that the $Z'$ only decays into the
three conventional fermion families.  If other decay channels were
possible, such as to exotic fermions filling out larger fermion
representations or supersymmetric partners, the $Z'$ width would be
larger, lowering the discovery limits.  On the other hand, if decays
to exotic fermions were kinematically allowed, the
discovery of exotic fermions would be an important discovery in
itself.   In addition, the $Z-Z'$ mixing is tightly constrained by
electro-weak precision measurements so we set it to zero
without affecting our conclusions.

The discovery limits for various models
at the colliders under discussion are summarized in fig. 1.
An important conclusion for these discovery limits is that these
bounds are relatively insensitive to specific models.  In addition,
since they are based on a distinct signal with little background
they are relatively robust limits.  For the case of the DI-TEVATRON
($\sqrt{s}=4$~TeV), the $p\bar{p}$ option has a 50\% higher discovery
reach than the $pp$ option
for a given luminosity indicating that valence quark contributions
to the Drell-Yan production process are still important at these
energies.

\subsection{$e^+e^-$ Colliders}

At $e^+e^-$ colliders discovery limits are indirect, being inferred
from deviations from the standard model predictions due to
interference between the $Z'$ propagator and the $\gamma$ and $Z^0$
propagators \cite{e+e-}.  This is similar to PEP/PETRA seeing the standard
model
$Z^0$ as deviations from the predictions of QED.
The basic process is $e^+e^- \to f\bar{f}$ where $f$
could be leptons $(e,\; \mu ,\; \tau)$ or quarks $(u, \; d, \; c,\;
s,\; b)$.  From the basic reactions a number of observables can be
used to search for the effects of $Z'$'s: The leptonic cross
section, $\sigma (e^+ e^- \to \mu^+
\mu^-)$, the ratio of the hadronic to the QED point cross section,
$R^{had}= \sigma^{had}/\sigma_0$, the leptonic forward-backward
asymmetry, $A^\ell_{FB}$, the leptonic  longitudinal asymmetry,
$A^\ell_{LR}$, the hadronic longitudinal asymmetry, $A^{had}_{LR}$,
the forward-backward asymmetry for specific quark or lepton
flavours, $A^f_{FB}$, the $\tau$ polarization asymmetry,
$A_{pol}^\tau$, and the polarized forward-backward asymmetry for
specific fermion flavours, $A^f_{FB}(pol)$. The indices $f=\ell, \;
q$, $\ell =(e,\mu,\tau)$, $q=(c, \; b)$,
and $had=$`sum over all hadrons' indicate the final state fermions.
The expressions for these observables are given in ref. \cite{capstick}.

For indirect limits, a 99\% C.L. corresponds to a $2\sigma$ effect of
one observable.
Since $2\sigma$ deviations are not uncommon one must be cautious
about how one obtains discovery limits for $Z'$'s.  One possibility
for obtaining believable limits is to raise the deviation required to
indicate the existence of a $Z'$.  A second possibility is to
combine several observables to obtain a $\chi^2$ figure of merit.
We follow the second approach here by including $\sigma^l$,
$R^{had}$, $A_{LR}$, and $A^{had}_{LR}$ to obtain the 99\% confidence
limits in fig. 1.\footnote{Although it is far from clear whether LEP200
will achieve any significant longitudinal polarization, $A_{LR}^{had}$
only  contributes significantly to the limit on $Z_\chi$ at LEP200 so
that our results are not in general sensitive to the inclusion of this
observable in the $\chi^2$ at LEP200.}

One sees that the discovery limits obtained at $e^+e^-$ colliders
are as large or larger than those that can be obtained at hadron
colliders.  However, the bounds obtained are more model dependent than
the  bounds obtained at hadron colliders.  For example, for model
$\psi$,
$C'_L=\pm C'_R$ so that either $C'_V$ or $C'_A=0$.  For
$\sqrt{s}$ sufficiently far away from the $Z^0$ pole deviations are
dominated by $Z^0-Z'$ and $\gamma-Z'$ interference which is proportional
to $C_V^2 C'_V^2 +2C_V C_A C'_V C'_A +C_A^2 C'_A^2 $.  Since for the
photon $C_A=0$, when $C'_V$ is also equal to 0 deviations from the
standard model become small.

Because
the bounds obtained at $e^+e^-$ colliders are indirect, based on deviations
from the standard model in precision measurements, they are sensitive to
the experimental errors, both statistical and systematic.  For example,
reducing the LEP200 integrated luminosity from 500~pb$^{-1}$ to
250~pb$^{-1}$ reduces the discovery limits by about 15\% and reducing
the NLC integrated luminosity from 50~fb$^{-1}$ to 10~fb$^{-1}$
(200~fb$^{-1}$ to 50~fb$^{-1}$) for the 500~GeV (1~TeV) case reduces the
discovery limit by about 33\%.  Including a 5\% systematic error in
cross section measurements and a 2\% systematic errors in asymmetries
where systematic errors partially cancel \cite{barklow} can lower the
discovery limits significantly.  The most extreme change is
for the sequential standard model $Z'$ which decreases by a factor of 2
at LEP200 and a factor of 3 at the NLC.  Clearly, systematic errors will
have to be kept under control for high precision measurements.

Finally,
we note that we did not include radiative corrections in our results.
In general this is an acceptable procedure since we are looking for
small deviations from the standard model predictions and radiative
corrections to $Z'$ contributions will be a small correction to a small
effect.  However, QED bremsstrahlung corrections, in particular, initial
state radiation, can give large contributions to the observables,
altering the statistics we assumed.  Since these are dependent on
details of the detector we have left them out but note that they can
alter the numerical values we show in fig. 1.

\section{CONCLUSIONS}

Among the facilities operating in the up-coming decade the TEVATRON
continues to raise  the limits on  new heavy gauge
bosons with  limits up to the 700-900 GeV range for ${\cal L}_{int}=1\
\hbox{fb}^{-1}$.  Depending on the luminosity, LEP200 can achieve
comparable limits for some of the models and could even surpass the
TEVATRON limits for the SSM and HARV models if the systematic errors can
be kept under control.

In the longer term, hadron colliders such as TEVATRON upgrades
and the LHC as well as the NLC  high luminosity $e^+e^-$ collider,
would significantly improve limits on the heavy gauge boson masses.
For typical models such limits are in the 1-2 TeV region for the
Tevatron upgrades, in the 4-5 TeV region for the LHC and roughly
$2-10\times \sqrt{s}$ for the NLC with 50~fb$^{-1}$.  The 60~TeV $pp$
LSGNA collider could achieve discovery limits up to 15~TeV or so while a
2~TeV could achieve limits ranging from 3~TeV for $Z_\psi$ to  20~TeV
for $Z_{HARV}$.  The limits obtained by hadron colliders are much less
model dependent than those obtained by $e^+e^-$ colliders.

The LHC and a high luminosity 500 GeV $e^+e^-$ collider have
discovery limits for a $Z'$ which are comparable.  However, limits
obtained from the LHC are robust, in the sense that they are
obtained from a direct measurement with little background.  On
the other hand, the limits obtained for the NLC are indirect, based
on statistical deviations from the standard model and are thus more
sensitive to having the systematic errors under control.

\acknowledgments

The author is most grateful to Tom Rizzo for many helpful
conversations and communications and to Mirjam Cveti\v{c} for her
encouragement to write this up.   This
research was supported in part by the Natural Sciences and Engineering
Research Council of Canada (NSERC).

\figure{Discovery limits for extra neutral gauge bosons ($Z'$) for the
models described in the text.  The discovery limits at
hadron collider are based
on 10 events in the $e^+e^-\ +\ \mu^+\mu^-$ channels while those for
$e^+e^-$ colliders are 99\% C.L. obtained from a $\chi^2$ based on
$\sigma (e^+e^- \to \mu^+\mu^-)$, $R^{had}=\sigma (e^+e^- \to
hadrons)/\sigma_0$, $A_{LR}^{\mu^+\mu^-}$, and $A^{had}_{LR}$.
The integrated luminosities are based on a $10^7$~sec year of running.}

\begin{table}
\caption{
Current constraints on $M_{Z'}$ (in GeV) for typical models from direct
production at the TEVATRON (${\cal L}_{int}=19.6$
pb$^{-1}$)\cite{cdf}, as well as
indirect limits from a global electro-weak analysis\cite{pdg}.  Both
sets of limits are at 95\% confidence level.}
\begin{tabular}{rccc}
model & direct & indirect \\
\hline
$\chi$ & $425 $ & $321$ \\
$\psi$ & 415  & 160  \\
$\eta$ & $440$ & $182$  \\
$LR$ & $445$ & $389$ \\
$SSM$ & 505 & 779 & \\
\end{tabular}
\end{table}


\begin{references}

\bibitem{haber}
H. Haber, {\sl Summary of the Electroweak Symmetry Breaking and Beyond
the Standard Model Working Group of the APS Study on the Future of
High Energy Physics}, Albaquerque New Mexico, August 7 1994.

\bibitem{capstick}
S. Capstick and S. Godfrey. Phys. Rev. {\bf D37}, 2466 (1988).

\bibitem{hewett}
J.L. Hewett and T.G. Rizzo,
{\sl Proceedings of the 1990 Summer Study on High Energy Physics}.
ed E. Berger, June 25-July 13, 1990, Snowmass Colorado
(World Scientific, Singapore, 1992) p. 222.

\bibitem{hewett2}
J.L. Hewett, {\sl Proceedings of the Workshop on Physics and
Experiments with Linear $e^+e^-$ Colliders},
ed. F.A. Harris, S.L. Olsen, S. Pakvasa, and X. Tata,
April 26-30, 1993, Waikoloa, Hawaii (World Scientific, Singapore, 1993)
p 246;
F. del Aguila, M. Cveti\v c, and P. Langacker, {\it ibid}, p. 490.

\bibitem{er5m}
J.L. Hewett and T.G. Rizzo, Phys. Rep. {\bf 183}, 193 (1989) and
references therein.

\bibitem{mohapatra}
For a review and original references see R.N. Mohapatra, {\sl
Unification and Supersymmetry} (Springer, New York, 1986).

\bibitem{chang}
D. Chang, R. Mohapatra, and M. Parida, Phys. Rev. {\bf D30}, 1052 (1984).

\bibitem{ma}
E. Ma, Phys. Rev. {\bf D36}, 274 (1987); K.S. Babu {\it et al.},
Phys. Rev. {\bf D36}, 878 (1987);
J.F. Gunion {\it et al.} Int. J. Mod. Phys. {\bf A2}, 118 (1987);
T. G. Rizzo, Phys. Lett. {\bf B206} 133 (1988).

\bibitem{georgi}
H. Georgi, E. Jenkins, and E.H. Simmons, Phys. Rev. Lett. {\bf 62}
2789 (1989); V. Barger and T.G. Rizzo, Phys. Rev. {\bf D41} 956
(1990).

\bibitem{othermodels}
A small sampling of other models with $Z'$'s is:
R. Foot and O. Hern\`andez, Phys. Rev. {\bf D41}, 946 (1990);
R. Foot, O. Hern\'andez, and T.G. Rizzo, Phys. Lett. {\bf B246}, 183 (1990);
A. Bagneid, T.K. Kuo, and N. Nakagawa, Int. J. Mod. Phys. {\bf A2}
1327 (1987); {\bf 2}, 1351 (1987)
R. Casalbuoni {it et al.}, Phys. Lett. {\bf B155}, 95 (1985);
Nucl. Phys. {\bf B310}, 181 (1988);
U. Baur {\it et al.}, Phys. Rev. {\bf D35}, 297 (1987);
M. Kuroda {\it et al.}, Nucl. Phys. {\bf B261}, 432 (1985);
K.T. Mahanthappa and P.K. Mohapatra, Phys. Rev. {\bf D42}, 1732
(1990); {\bf 42}, 2400 (1990).

\bibitem{langacker}
P. Langacker and M. Luo, Phys. Rev. {\bf D45}, 278 (1992).

\bibitem{ew}
G. Altarelli, R. Barbieri, and S. Jadach, Nucl. Phys. {\bf B369}, 3 (1992);
G. Altarelli {\sl et al}., Phys. Lett {\bf B263}, 459 (1991);
Phys. Lett. {\bf B261}, 146 (1991); {\bf 245}, 669 (1990); Nucl.
Phys. {\bf B342}, 15 (1990);
Amaldi {\it et al.}, Phys. Rev. {\bf D36}, 1385 (1987);
L.S. Durkin and P. Langacker, Phys. Lett {\bf B166}, 436 (1986);
F.M. Renard and C. Verzegnassi, Phys. Lett. {\bf 260}, 225 (1991);
F. del Aguila, J.M. Moreno, and M. Quiros, Nucl. Phys., {\bf B361}, 45 (1991);
F. del Aguila, W. Hollik, J.M. Moreno, and M. Quiros, Nucl. Phys.,
{\bf B372}, 3 (1992);
Phys. Lett. {\bf B254}, 479 (1991); Phys. Rev. {\bf D40}, 2481 (1989);
M.C. Gonzalez-Garcia and J.W. F. Valle, Nucl. Phys. {\bf B345}, 312
(1990); Phys. Lett. {\bf B236} 360 (1990); {\bf 259}, 365 (1991);
A. Djouadi {\it et al.}, Nucl. Phys. {\bf B349}, 48 (1991);
V. Barger, J.L. Hewett and T.G. Rizzo, Phys. Rev. {\bf D42}, 152 (1990);
T.G. Rizzo, Phys. Rev. {\bf D40}, 3035 (1989);
T.G. Rizzo, {\sl Proceedings of the 1990 Summer Study on High Energy
Physics}. ed E. Berger, June 25-July 13, 1990, Snowmass Colorado
(World Scientific, Singapore, 1992) p. 233.

\bibitem{pdg}
Particle Data Group, Phys. Rev. {\bf D50}, 1173 (1994).

\bibitem{cdf}
CDF Collaboration, Abe {\it et al.}, FERMILAB-PUB-94-198-E (July, 1994).

\bibitem{pp}
P. Langacker, R.W. Robinett, and J.L. Rosner,
Phys. Rev. {\bf D30}, 1470 (1984);
F. del'Aguila, J.M. Morena, and M. Quiros, Phys. Rev. {\bf D40}, 2481 (1989);
T.G. Rizzo, Phys. Rev. {\bf D48}, 44705 (1993);
J.L. Rosner, Phys. Rev. {\bf D35}, 2244 (1987);
V. Barger {\it et al.}, Phys. Rev. {\bf D35}, 2893 (1987);
F. del'Aguila, M. Quiros, and F. Zwirner, Nucl. Phys. {\bf B287}, 419 (1987);
V. Barger, N.G. Deshpande, and K. Whisnant, Phys. Rev. {\bf D35}, 1005 (1987).

\bibitem{ehlq}
E. Eichten, I. Hinchliffe, K.D. Lane, and C. Quigg, Rev. Mod. Phys.
{\bf 56}, 579 (1984).

\bibitem{rizzo}
J.L. Hewett and T.G. Rizzo, Phys. Rev. {\bf D45}, 161 (1992).

\bibitem{e+e-}
F. Boudjema, B.W. Lynn, F.M. Renard, C. Verzegnassi,
Z. Phys. {\bf C48}, 595 (1990);
A. Blondel, F.M. Renard, P. Taxil, and C. Verzegnassi, Nucl. Phys.
{\bf B331}, 293 (1990);
G. Belanger and S. Godfrey, Phys. Rev. {\bf D34}, 1309 (1986);
{\bf D35}, 378 (1987);
P.J. Franzini and F.J. Gilman, Phys. Rev. {\bf D35}, 855 (1987);
M. Cveti\v{c} and B. Lynn, Phys. Rev. {\bf D35}, 1 (1987);
B.W. Lynn and C. Verzegnassi, Phys. Rev. {\bf D35}, 3326 (1987);
T.G. Rizzo {\it ibid}, {\bf 36}, 713(1987);
A. Bagneid, T.K. Kuo, and G.T. Park {\it ibid}, {\bf 44}, 2188 (1991);
A. Djouadi, A. Leike, T. Riemann, D. Schaile and C. Verzegnassi,
Z. Phys. {\bf C56} 289 (1992);
A. Leike, Z. Phys. {\bf C62}, 265 (1994);
J.L. Hewett and T.G. Rizzo, {\sl  Proceedings of Physics and
Experiments with Linear COlliders}, ed. R. Orava {\it et al.},
Saariselk\"a, Finland (World Scientific, 1992); T.G. Rizzo, {\it ibid}.

\cite{barklow} These systematic errors are based on simulations of an
SLD type detector operating at a 500~GeV $e^+e^-$ collider.  T. Barklow,
private communication.

\end{references}
\end{document}